\documentclass[letterpaper]{article}
\usepackage{latexsym}
\usepackage{graphicx}
\begin{document}

\def\beq{\begin{equation}}
\def\eeq{\end{equation}}
\def\fdotf{\overrightarrow{f} \cdot \overrightarrow{f}}
\def\ttilde{\tilde{t}}
\def\1fdotf{\frac{1}{\left(1+\fdotf\right)}}
\def\e{\epsilon}
\def\d{\delta}
\def\p{\partial}
\def\O{\mathcal{O}}
\def\F{\mathcal{F}}
\def\X{\underline{\overline{X}}}
\def\ie{\emph{i.e.}}
\def\psibar{\overline{\psi}}
\def\thetabar{\overline{\theta}}
\def\vare{\varepsilon}
\def\rhobar{\overline{\rho}}
\def\fvector{\overrightarrow{f}}
\def\half{\frac{1}{2}}
\def\afourth{\frac{1}{4}}

\title{%
\bf The Parisi-Sourlas Mechanism in Yang-Mills Theory?
}
\author{%
Jose A. Magpantay%
\footnote{Alexander von Humboldt fellow; on sabbatical leave from the
University of the Philippines} %
\footnote{\it jamag@thep.physik.uni.mainz.de, 
jamag@nip.upd.edu.ph}\\
Institut f$\ddot{u}$r Physik%
\footnote{Address until 30 Oct 1999},
Universit$\ddot{a}$t Mainz\\
$55009$ Mainz, Germany\\
and\\
Research Center for Theoretical Physics\\
Jagna, Bohol, Philippines
}
\date{\today}

\maketitle

\begin{abstract}
The Parisi-Sourlas mechanism is exhibited in pure Yang-Mills theory.
Using the new scalar degrees of freedom derived from the non-linear gauge
condition, we show that the non-perturbative sector of Yang-Mills theory
is equivalent to a $4$D $O(1,3)$ sigma model in a random field.
We then show that the leading term of this equivalent theory is invariant
under supersymmetry transformations where \(x^{2}+\thetabar\theta\) is
unchanged.
This leads to dimensional reduction proving the equivalence of the
non-perturbative sector of Yang-Mills theory to a $2$D $O(1,3)$ sigma 
model.
\end{abstract}

\section{Introduction}

$\;\;\;\;$
There are now several new ideas on how to confine quarks
inside hadrons in recent literature.
The Seiberg-Witten paper \cite{Seiberg-Witten} on the spontaneous breaking
of N=$2$ supersymmetric Yang-Mills (SYM) to N=$1$ SYM is supposed to
implement the dual type II superconductivity model of confinement proposed
by Mandelstam and `t Hooft \cite{'tHooft}.
Last year, Fadeev and Niemi \cite{Fad-Niemi} derived an effective field
theory for knotlike solitons that is supposed to describe the infrared
regime of pure Yang-Mills theory.
In both approaches, confinement is essentially achieved through
chromoelectric flux tubes, which provide the linear potential.
These mechanisms break local Lorentz invariance because of the preferred
direction defined by the flux tube.

There is a third approach to confinement, which is based on the non-linear
generalization of the Coulomb gauge \cite{Jamag:MPLa}.
In a series of papers \cite{Jamag:PTP,Jamag:etc,Jamag:workshop}, this
author argued the importance of generalizing the Coulomb gauge to the gauge
condition
\beq
\label{gauge condition}
 (\partial \, \cdot\, D)(\partial \, \cdot \, A) = 0,
\eeq
because field configurations which satisfy \(\partial \, \cdot \,  A^{a}
= f^{a} \neq 0 \), and \( (\partial \, \cdot \, D) f = 0 \), are not
gauge transformable to the Coulomb surface.
In the non-linear regime (the linear sector being the Coulomb gauge) of the
gauge condition (\ref{gauge condition}), the Yang-Mills potential can be
written as
\beq
\label{YMpotential}
 A^{a}_{\mu} = \frac{1}{\left(1+\fdotf\right)}
 (\delta^{ab} + \e^{abc} f^{c} + f^{a} f^{b})
 (\frac{1}{g}\partial_{\mu} f^{b} + t^{b}_{\mu}).
\eeq
As discussed in reference \cite{Jamag:workshop}, the effective dynamics of
the new scalar field $f^{a}$ hints at non-perturbative effects.
This was confirmed in reference \cite{Jamag:MPLa} where the linear behavior
of the instantaneous gluon propagator and the area law behavior of the
Wilson loop were explicitly shown.
The mechanism for these behaviors is not quantum mechanical but simply a
statistical treatment of classical solutions of the action for pure $f$'s
(all spherically symmetric $f^{a}(x)$ in $4$D Euclidean).
This confinement mechanism then maintains local Lorentz invariance unlike
the mechanisms that make use of chromoelectric flux tubes.

Aside from not breaking Lorentz invariance, this mechanism relies on a
straightforward, though non-trivial, difference between Abelian and
non-Abelian theories.
In an Abelian theory, gauge transformation is only a ``translation'', thus
all orbits pass through the Coulomb surface.
Furthermore, the Coulomb gauge describes the physical degrees of freedom,
the transverse photon.
But in a non-Abelian theory, gauge transformation is a combination of
``translation'' and ``rotation'' that depend on the gauge field being
transformed.
This fact leads to complications like gauge copying phenomena \cite{Gribov}
and the existence of gauge fields that cannot be gauge transformed to the
Coulomb gauge \cite{Jamag:etc,Jamag:workshop}. Hence, the proposal to
generalize the Coulomb gauge to the non-linear gauge defined by equation
\ref{gauge condition}.
Note that in the Abelian limit or in the short distance limit (weak
coupling), the non-linear gauge reduces to the Coulomb gauge (\(f=0\)).
But even before the coupling becomes very large, the fact that the
non-linear term is proportional to $p^{2}$ ($p$ = momentum) while the linear
term depends on $p^{3}$, the non-linear term would increasingly be more
important in the long-distance regime.
The non-zero $f$ terms must now be taken into account and their dynamics
provide the confinement mechanism.

In this paper, we will prove a conjecture made in references
\cite{Jamag:MPLa} and \cite{Jamag:workshop} on the implementation of the
Parisi-Sourlas mechanism \cite{Parisi-Sourlas} in pure Yang-Mills theory.
The conjecture is based on two facts.
First, the action for the pure $f$ term is clearly non-perturbative because
of ($\frac{1}{g^{2}}$) and it is infinitely non-linear.
Second, the action for $f$ is proportional to \((\partial f)^{4}\), thus
hinting that it can be written as \((\frac{\delta S}{\delta f})^{2}\), where
$S$ is an action with a usual kinetic term (\(\sim (\partial f)^{2}\)).
If indeed we can write the action for the $f$ term as
\((\frac{\delta S}{\delta f})^{2}\), then its dynamics is stochastic,
{\it i.e.}, driven by a random field.
Clearly this hints of the possible realization of the Parisi-Sourlas
mechanism.

The outline of this paper is as follows.
In section \ref{section:II}, we give a detailed derivation of the
path-integral in the non-linear gauge.
Section \ref{section:III} will prove the equivalence of the non-perturbative
regime of Yang-Mills to an $O(1,3)$ sigma model in random field.
Section \ref{section:IV} presents the proof of dimensional reduction.
The conclusion summarizes what the paper achieves.

\section{The path-integral in the non-linear gauge}
\label{section:II}

$\;\;\;\;$
The Yang-Mills path-integral in the gauge defined by equation
\ref{gauge condition} is
\beq
\label{eq:3}
 PI = \int (dA^{a}_{\mu})\; \delta(\partial \, \cdot \, D(\partial \cdot
  A)) \; det \mathcal{O}e^{-S_{YM}(A)},
\eeq
where $\mathcal{O}$ is the Fadeev-Popov operator
\beq
\label{eq:4}
 \mathcal{O}^{ad} = (D \, \cdot \, \partial)^{ab} (\partial \, \cdot \, 
  D)^{bd} - \epsilon^{abc} ( \partial f^{b}) \, \cdot \, D^{cd}.
\eeq
The important point about $\mathcal{O}$ is that it is a non-singular
operator even though \(\partial \, \cdot \, D\) is singular
\cite{Jamag:workshop}.
The reason for this is that the gauge transform of a field configuration
that satisfies \(\p\cdot A^{a} = f^{a}\),
and \((\p\cdot D)^{ab}f^{b}=0\), remains on the surface \(\p\cdot
A^{a}=f^{a}\) but no longer in the horizon.
This can be seen by considering an infinitesimal parameter \(\Lambda^{a} 
= \e f^{a}\), resulting in an \(A'^{a}_{\mu} = A^{a}_{\mu} + \e
D^{ab}_{\mu} f^{b}\), thus satisfying \(\p\cdot A'^{a} = f^{a}\), but with 
\((\p\cdot D^{ab}(A'))(\p\cdot A'^{b}) =  -\e\e^{abc} D^{cd}_{\mu}
f^{d}\p_{\mu}f^{b} \neq 0\).
This should rectify the inconsistent claim in reference (7) that the
field configurations in the non-linear sector of the non-linear
gauge never leave the horizon yet having a non-singular $\mathcal{O}$.

Let us introduce the scalar fields by inserting the following identity in
equation \ref{eq:3}:
\beq
\label{eq:5}
 \mathcal{I} = \int (d f^{a}) \delta (f^{a} - \partial \, \cdot \, A^{a}).
\eeq
The two delta functionals (see equations \ref{eq:5} and \ref{eq:3}) imply
that we can write
\beq
\label{eq:6}
 l^{2}_{0} D^{ab}_{\mu} f^{b}
  = (A^{a}_{\mu} - \partial_{\mu} \frac{1}{\Box^{2}} f^{a}) 
    - t^{a}_{\mu},
\eeq
where $t^{a}_{\mu}$ is a transverse vector field and $l_{0}$ is a length
scale introduced for dimensional reasons.
Equation \ref{eq:6} can be solved for $A^{a}_{\mu}$ in terms of $f^{a}$ and
$t^{a}_{\mu}$:
\beq
\label{eq:7}
 A^{a}_{\mu} = \frac{1}{\left(1 + g^{2} l^{4}_{0} \fdotf\right)}\;
  (\delta^{ab} + \epsilon^{abc} gl^{2}_{0} f^{c} + g^{2}l^{4}_{0} f^{a} f^{b})
  (l^{2}_{0} \partial_{\mu} f^{b} + t^{b}_{\mu}).
\eeq
Shifting \(t^{a}_{\mu}+\partial_{\mu} \frac{1}{\Box^{2}}f^{b}
\longrightarrow t^{a}_{\mu}\) and rescaling \(gl^{2}_{0}f^{a}
\longrightarrow f^{a}\), we find the expression given by equation
\ref{YMpotential}.
The new vector field $t^{a}_{\mu}$ and the original potential $A^{a}_{\mu}$
will now have divergence equal to \(\frac{1}{gl^{2}_{0}} f^{a}\).

What equation \ref{YMpotential} means is that we traded the \(12
A^{a}_{\mu}\)'s satisfying the three constraints given by equation
\ref{gauge condition} to the \(12\; t^{a}_{\mu}\)'s and \(3\; f^{a}\)'s.
The $t^{a}_{\mu}$'s satisfy \(\partial_{\mu} t^{a}_{\mu} =
\frac{1}{gl^{2}_{0}} f^{a}\) ($3$ conditions).
The extra degrees of freedom are removed by the constraints
\begin{eqnarray}
 \rho^{a} & = &
  \frac{1}{\left(1+\fdotf\right)^{2}}
  [ \epsilon^{abc} + \epsilon^{abd} f^{d}f^{c} - \epsilon^{acd} f^{d} f^{b} 
  + f^{a} f^{d} \epsilon^{dbc}
  \nonumber \\[-2mm]%
  &  & \label{eq:8} \\[-2mm]%
  & - &
  f^{a} (1+\fdotf) \delta^{bc}
  - f^{c} (1 + \fdotf) \delta^{ab}]
  \partial_{\mu}f^{b}t^{c}_{\mu} = 0,
  \nonumber 
\end{eqnarray}
which are to guarantee \(\partial^{\mu}A^{a}_{\mu} = \frac{1}{gl^{2}_{0}}
f^{a}\).

Let us incorporate $t^{a}_{\mu}$ in the path-integral by inserting the
identity:
\beq
\label{eq:9}
 \mathcal{I} = \int (d t^{a}_{\mu}) \: \delta (t^{a}_{\mu} -
\ttilde^{a}_{\mu}),
\eeq
where \(\ttilde^{a}_{\mu}\) is solved from equation
(\ref{YMpotential}).
This means
\beq
\label{eq:10a}
 \delta(A^{a}_{\mu} - \1fdotf (\delta^{ab} + \e^{abc} f^{c}
  + f^{a} f^{b})(\frac{1}{g} \partial_{\mu} f^{b} + t^{b}_{\mu}))
  = \frac{\delta (t^{a} - \ttilde^{a}_{\mu})}{det\left(
  \frac{\delta A^{a}_{\mu}(x)}{\delta t^{b}_{\nu}(x)}\right)},
\eeq
where
\beq
\label{eq:10b}
 \frac{\d A^{a}_{\mu}(x)}{\d t^{b}_{\nu}(x')}
  = \mathcal{F}^{ab} \d_{\mu \nu} \d^{4}(x-x'),
\eeq
\beq
\label{eq:10c}
\mathcal{F}^{ab} = \1fdotf (\d^{ab} + \e^{abc} f^{c} + f^{a} f^{b}).
\eeq
The path-integral now becomes
\[
 PI = \int (dt^{a}_{\mu})(df^{a})(dA^{a}_{\mu}) \d (\p \cdot D(\p \cdot A))
  \d (\p \cdot A^{a} - \frac{1}{gl^{2}_{0}} f^{a})
\] 
\beq
\label{eq:11}
  \times \d (A^{a}_{\mu} - \mathcal{F}^{ab} (\frac{1}{g} \p_{\mu}
  f^{b}+t^{b}_{\mu}))
  det (\mathcal{O}) det \left( \frac{\d A^{a}_{\mu}}{\d
   t^{b}_{\nu}}\right) e^{-S_{YM}(A)}.
\eeq
Before integrating out $A^{a}_{\mu}$, we can change the set of
constraints from
\beq
\label{eq:12a}
 \phi^{a} = (\p \cdot D)^{ab} (\p \cdot A^{b}) = 0,
\eeq
\beq
\label{eq:12b}
 \chi^{a} = \p \cdot A^{a} - \frac{1}{gl^{2}_{0}} f^{a} = 0,
\eeq
to
\beq
\label{eq:13a}
 \phi'^{a} = \rho^{a},
\eeq
\beq
\label{eq:13b}
 \chi'^{a} = \p \cdot t^{a} - \frac{1}{gl^{2}_{0}} f^{a}.
\eeq
This is implemented by
\beq
\label{eq:14a}
 \d (\phi^{a}) \d (\chi^{b})
  = \d (\phi'^{a}) \d (\chi'^{b}) J^{-1}\left(
   \frac{\phi,\chi}{\phi',\chi'}
  \right)
\eeq
where
\beq
\label{eq:14b}
 J (\frac{\phi,\chi}{\phi',\chi'}) = det\left(
 \left[
 \begin{array}{cc}
 \d^{ab} & \mathcal{F}^{ab}-\d^{ab} \\
 \d^{ab} & \mathcal{F}^{ab}
 \end{array}
 \right]
 \d^{4}(x-x')
 \right)
\eeq
This Jacobian is $1$, which can be shown by direct evaluation or by
introducing fermionic coordinates and then doing the integrations.
Substituting the new constraints, evaluating the determinant given by
equations (\ref{eq:10b},\ref{eq:10c}), and integrating out $A^{a}_{\mu}$, we
find
\beq
\label{eq:15}
 PI = \int (dt^{a}_{\mu})(df^{a})\: \d (\p \cdot t^{a} -%
  \frac{1}{gl^{2}_{0}}%
  f^{a})\: \d (\rho^{a})\: det^{-4}(1+\fdotf)\:%
  det \mathcal{O} e^{-S_{YM}},
\eeq
where it is understood that equation (\ref{YMpotential}) is substituted in
$\mathcal{O}$ and $S_{YM}$.

The Yang-Mills field strength in terms of $f^{a}$ and $t^{a}_{\mu}$ has the
following form
\beq
\label{eq:16}
 F^{a}_{\mu \nu} = \frac{1}{g} Z^{a}_{\mu \nu}%
  + L^{a}_{\mu \nu} + g Q^{a}_{\mu \nu}(f;t),
\eeq
where $Z^{a}_{\mu\nu}$, $L^{a}_{\mu \nu}$, and $gQ^{a}_{\mu \nu}(f;t)$ are
zeroth-order linear and quadratic in $t^{a}_{\mu \nu}$, respectively.
The action becomes
\begin{eqnarray}
 S_{YM} & = & \int \left[ \frac{1}{g^{2}}\: Z^{a}_{\mu \nu}(f)Z^{a}_{\mu%
  \nu}(f)%
  + \frac{2}{g}Z^{a}_{\mu \nu}(f)L^{a}_{\mu \nu}(f;t)
  \right. \nonumber \\%
  & + &
  \left.
  \left[ 2 Z^{a}_{\mu \nu}(f) Q^{a}_{\mu \nu}(f;t)%
  + L^{a}_{\mu \nu}(f;t)L^{a}_{\mu \nu}(f;t) \right]
  \right.
  \label{eq:17} \\%
  & + &
  \left.
  2 g L^{a}_{\mu \nu}(f;t)Q^{a}_{\mu \nu}(f;t)%
  + g^{2} Q^{a}_{\mu\nu}(f;t) Q^{a}_{\mu\nu}(f;t)\right].\nonumber
\end{eqnarray}

\section{Equivalence to a non-linear sigma model}
\label{section:III}

$\;\;\;\;$
Consider the pure $f$ path-integral
\beq
\label{eq:18}
 PI(f) = \int (df^{a})\: det^{-1}(1+\fdotf)\: det \widetilde{\mathcal{O}} \:%
  e^{-\frac{1}{g^{2}}\int d^{4} \frac{1}{4} Z^{2}},
\eeq
where $\O$ is the operator given by equation (\ref{eq:4}) with
\(t^{a}_{\mu} = 0\).
To get the minus one power of \(det(1 + \fdotf)\), we changed the
$(1+\fdotf)$ factor of $\rho^{a}$ from minus two to plus one (see equation
(\ref{eq:8})).
Why this change was made will be clarified later.
In isolating the path-integral for pure $f$ term (as given in equation
(\ref{eq:18})) from the full path-integral, we are essentially claiming that
the pure $f$ dynamics is dominant and the remaining $(f^{a};t^{a}_{\mu})$
constrained dynamics can be treated as corrections.

We will now show that this path-integral is equal to
\beq
\label{eq:19a}
 PI(f) = \int (df^{a})\: det\left(\frac{\d^{2}S}{\d f^{a} \d f^{b}}\right)\:
  exp\left[ -\frac{1}{2} \int d^{4}x \left( \frac{\d S}{\d%
  f^{a}}\right)^{2} \right],
\eeq
where
\beq
\label{eq:19b}
 S = \frac{1}{2g} \int d^{4}x\; \eta^{ab} \p_{\mu} f^{a} \p_{\mu} f^{b},
\eeq
\beq
\label{eq:19c}
 \eta^{ab} = -\d_{ab} + \frac{f^{a} f^{b}}{(1+\fdotf)}.
\eeq
Equations (\ref{eq:19b},\ref{eq:19c}) give the action for an $O(1,3)$ sigma
model in the non-liner form.
Equation (\ref{eq:19a}) says that the non-perturbative regime of Yang-Mills
theory is equivalent to an $O(1,3)$ sigma model in a random field.
The rest of this section is devoted to the proof of this claim.

The pure $f^{a}$ field strength can be written as
\beq
\label{eq:20a}
 Z^{a}_{\mu \nu} = \X^{abc} \: \p_{\mu} f^{b}\: \p_{\nu}f^{c},
\eeq
\begin{eqnarray}
 \X^{abc} & = &
 \frac{1}{(1+\fdotf)^{2}} \: %
  \left\{
   (1+2\fdotf)\e^{abc} + 2\d^{ab}f^{c} - 2\d^{ac}f^{b}%
   \right. \nonumber \\[-2mm]
   &   & \label{eq:20b}  \\[-2mm]
   & + &
   \left.
   3\e^{abd}f^{d}f^{c} - 3\e^{acd}f^{db} + \e^{bcd}f^{a}f^{d}
  \right\}. \nonumber
\end{eqnarray}
Equations (\ref{eq:20a},\ref{eq:20b}) give
\begin{eqnarray}
 \frac{1}{4}Z^{2} & = &
  (\frac{1}{4})\: \frac{1}{(1+\fdotf)^{2}}%
  \left\{%
   (1+\fdotf)^{2}%
   \left[(\p_{\mu}f^{a}\:\p_{\mu}f^{a})^{2}%
    -(\p_{\mu}f^{a}\: \p_{\nu}f^{a})^{2}
   \right]
 \right. \nonumber \\[-2mm]%
 &   & \label{eq:21}  \\[-2mm]%
 & + &
 \left.
  6(1+2\fdotf)%
  \left[
   (\p_{\mu}f^{a})^{2} (f^{b}\p_{\nu}f^{b})^{2}%
   - (\p_{\mu}f^{a}f^{b}\p_{\mu}f^{b})^{2}
  \right]
  \right\}. \nonumber
\end{eqnarray}
It is interesting to note that if we impose $\fdotf=1$, equation
(\ref{eq:21}) gives the interaction term of the Fadeev-Niemi action,
which supposedly describes knotted strings \cite{Fad-Niemi}. This
observation shows that indeed the $f^{a}$ dynamics describes the 
non-perturbative regime of Yang-Mills theory.

The field equation of the non-linear sigma-model is:
\beq
\label{eq:22a}
\frac{\d S}{\d f^{a}} =
\eta^{ab}\Box^{2}f^{b} - \frac{f^{a}}{(1+\fdotf)}\:\eta^{bc}\p_{\mu}f^{b}
\p_{\mu} f^{c}.
\eeq
If we identify $f^{a}$ of above with the $f^{a}$ of Yang-Mills in the 
non-linear gauge, then we should have
\begin{eqnarray}
\Box^{2}f^{a} & = & \frac{1}{1+\fdotf}
 \left[
  \p_{\mu}f^{a}f^{b}\p_{\mu}f^{b} - f^{a}\p_{\mu}f^{b}\p_{\mu}f^{b}
  + \e^{abc}f^{c}\p_{\mu}f^{b}f^{d}\p_{\mu}f^{d}
 \right]
 \nonumber \\[-1mm]
\label{eq:22b} & & \\[-1mm]
 & &
  + {\rm\;linear\;term\;in}\; t^{a}_{\mu}
 \nonumber
\end{eqnarray}
Equation \ref{eq:22b} comes from the non-linear gauge condition. Since we
are considering pure $f$ dynamics, we will put $t^{a}_{\mu}=0$.
Clearly we find that
\[
\frac{1}{4}\int d^{4}x Z^{2} \neq \frac{1}{2} \int d^{4}x
 \left(
  \frac{\d S}{\d f^{a}}
 \right)^{2}.
\]
However, we can add a zero, a surface term, to the Yang-Mills action such
that
\beq
\label{eq:23}
 \frac{1}{4}\int d^{4}x Z^{2} + \int d^{4}x \p_{\mu}H_{\mu} =%
  \frac{1}{2}\int d^{4}x%
   \left(
    \frac{\d S}{\d f^{a}}
   \right)^{2}.
\eeq
Inspection shows that
\begin{eqnarray}
 \p_{\mu}H_{\mu} & = &
 \alpha (\p_{\mu}f^{a}\: \p_{\mu}f^{a})^{2}%
 + \beta (\p_{\mu}f^{a}\: \p_{\nu}f^{a})^{2}%
 \nonumber \\%
 & + &
 \left.
 \gamma (\p_{\mu}f^{a})^{2} (f^{b}\p_{\nu}f^{b})^{2}%
 + \d (\p_{\mu}f^{a}f^{b}\p_{\mu}f^{b})^{2}%
 \right.  \label{eq:24} \\%
 & +  &
 \rho \left[%
 (f^{a}\p_{\mu}f^{a})^{2}
 \right]^{2}, \nonumber
\end{eqnarray}
and equating corresponding terms in (\ref{eq:23}) yields
\begin{eqnarray}
 \alpha & = &
 \frac{2(\fdotf)^{3} + 7(\fdotf)^{2} - 6(\fdotf) -1}%
  {4(1+\fdotf)^{4}},
 \nonumber \\%
 \beta & = &
 \frac{1}{4(1+\fdotf)^{2}},
 \nonumber \\%
 \gamma & = &
 \frac{(\fdotf)^{3} - 11(\fdotf) - 7}%
 {2(1+\fdotf)^{4}},
 \nonumber \\%
 \d & = & 
 \frac{(\fdotf)^{2} + 8(\fdotf) + 4}%
 {2(1+\fdotf)^{4}},%
 \nonumber \\%
 \rho & = &
 \frac{-%
 \left[%
  (\fdotf)^{2} + 3(\fdotf) + 1%
 \right]}%
 {2(1+\fdotf)^{4}}. \nonumber
\end{eqnarray}

To prove that the added term is indeed zero, we note that in $4$D Euclidean
space,
\[
 H_{\mu}(x) = \p^{x}_{\mu} \; \int d^{4}y \:%
  \frac{1}{(x-y)^{2}}\: \left[\p \cdot H\right]_{y},
\]
giving
\beq
\label{eq:25}
 \int d^{4}x\: \p_{\mu}H_{\mu} =%
  \oint_{x\rightarrow\infty}%
  \left\{
   \int d^{4}y\: \frac{(x-y)_{\mu}}{(x-y)^{3}}
   \left[
    \p \cdot H
   \right]_{y}
  \right\} n_{\mu}dS.
\eeq
Since we will only be considering $L^{2}$ fields, \ie,
\(\int d^{4}x A^{a}_{\mu}A^{a}_{\mu} =\) finite, then
\(A^{a}_{\mu} \sim \frac{1}{x^{2 + \e}}\) as the $4$D Euclidean radius
$x\rightarrow\infty$.
From equation (\ref{YMpotential}), we must have
\(f \sim \frac{1}{x^{1+\e}}\) as $x\rightarrow\infty$.
Shifting coordinates in equation (\ref{eq:25}), \ie,
\(x_{\mu}-y_{\mu}=y'_{\mu}\), we find
\begin{eqnarray}
 \int d^{4}x\: (\p \cdot H) & = &
 \oint_{x\rightarrow\infty}%
 \left\{
 \int d^{4}y' \frac{y'_{\mu}}{y'^{3}}\: %
  \left[
   (\p\cdot H)_{x} - \p_{\alpha}(\p\cdot H)_{x}y'_{\alpha}%
  \right.%
 \right. \nonumber \\[-2mm]%
 &   &   \label{eq:26}   \\[-2mm]%
 & + &
 \left.%
  \left.%
  \frac{1}{2}\p_{\alpha}\p_{\beta}(\p\cdot H)_{x}y'_{\alpha}y'_{\beta}%
  \cdots
  \right]
 \right\} n_{\mu}dS. \nonumber
\end{eqnarray}
We only have to look at the even number of integrands in $y'$ because all the
odd vanish.
From the expression of $(\p\cdot H)$, we see that
\[
 (\p\cdot H) \sim%
 \left\{%
  \frac{1}{x^{16+\e}},\; \frac{1}{x^{14+\e}},\; \frac{1}{x^{12+\e}},\;%
  \frac{1}{x^{10+\e}}
 \right\}_{x\rightarrow\infty}.
\]
Because of these behaviors, even though each $y'$ integration diverges as
\(y'\rightarrow\infty\), the $H$ factor that goes with it goes to zero
faster as $x\rightarrow\infty$ resulting in a zero for each term.
Thus,
\beq
\label{eq:27}
 \int d^{4}x\: \p_{\mu}H_{\mu} = \oint_{x\rightarrow\infty} H_{\mu}\:%
 n_{\mu}dS = 0.
\eeq

To complete the proof of the equivalence to an $O(1,3)$ sigma model in
random field, we need to establish the following determinant relation
\beq
\label{eq:28}
 det^{-1}(1+\fdotf)\: det\widetilde{\O} \approx det%
 \left(%
  \frac{\d^{2}S}{\d f^{a} \d f^{b}}
 \right).
\eeq
The proportionality constant may be infinite but it should be field
$f$ independent.
At first glance, equation (\ref{eq:28}) does not seem to make sense because
$\O$ is a fourth-order differential operator while \(\frac{\d^{2}S}{\d f\d 
f}\) is only second-order.
However, as we will show in the following arguments equation (\ref{eq:28}) 
is valid.

It is important to note that we can write
\beq
\label{eq:29a}
 \frac{\d^{2}S}{\d f\d f} = - \eta (\Box^{2} + R(f) + M_{\mu}(f)\p_{\mu}),
\eeq
where 
\beq
\label{eq:29b}
 M^{ab}_{\mu} = - 2 f^{a}\p_{\mu} f^{b} + \frac{
 2 f^{a}f^{b}f^{c}\p_{\mu}f^{c}}
 {1+\fdotf},
\eeq
\begin{eqnarray}
 R^{ab} & = & \frac{\left[(2+\fdotf)f^{a}f^{b}-\d^{ab}
  \right]} {\left(1+\fdotf\right)^{2}}
  \;\p_{\mu}f^{c}\p_{\mu}f^{c}
  \nonumber \\[-1mm]
& + & \frac{f^{a}\p_{\mu}f^{b}f^{c}\p_{\mu}f^{c}}
 {1+\fdotf}
 - \frac{2 f^{a}f^{b}f^{c}\p_{\mu}f^{c}f^{d}\p_{\mu}f^{d}}
 {\left(1+\fdotf\right)^{2}} 
\label{eq:29c} \\[-1mm]
& - & \frac{f^{a}\e^{bcd}f^{d}\p_{\mu}f^{c}f^{e}\p_{\mu}f^{e}}
 {\left(1+\fdotf\right)}
\nonumber,
\end{eqnarray}
and $\eta$ is the ``metric'' given by equation (\ref{eq:19c}).
The factoring out of $\eta$ is important for two reasons.
First, \(det\:\eta = det^{-1}(1+\fdotf) = det\:\F\), and this cancels out
the extra determinant at the LHS of equation (\ref{eq:28}).
Second, it leaves $\Box^{2}$ without a field-dependent coefficient,
which in turn enables us to carry out the following identification:
\beq
\label{eq:30}
det\:\tilde{\mathcal{O}} = det\:\Box^{2}\;det(\Box^{2}+M\cdot\p + R).
\eeq
Equation (\ref{eq:30}) implies that the proportionality
factor in equation (\ref{eq:28}) is $det\;\Box^{2}$.
It is hinted by the fact that for 
\(f^{a}={\rm constant}\), 
\(\tilde{\mathcal{O}}=(\Box^{2})^{2}\) and \(R=M_{\mu}=0\).

Now we prove that equation (\ref{eq:30}) is true for any $f^{a}$.
Neglecting the $t$ dependent term in equation (\ref{eq:22b}), we find
\(\p_{\mu}\tilde{A}^{a}_{\mu}=0\).
This does not mean, however, that we are expanding about a field
configuration on the Coulomb surface.
Equation (\ref{YMpotential}) clearly shows that
this is not a case of background decomposition (where \(A = \tilde{A}+a\))
because $t^{a}_{\mu}$ is linked in a very nonlinear manner to $f^{a}$.
Besides, we are not considering a particular field configuration $f^{a}$.
Thus, if anything, the vanishing of the divergence of
\(\tilde{A}^{a}_{\mu}\) is purely a coincidence, but a welcome one because
it leads to a significant amplification of $\mathcal{O}$.

\begin{figure}
\centering
\includegraphics[totalheight=2in]{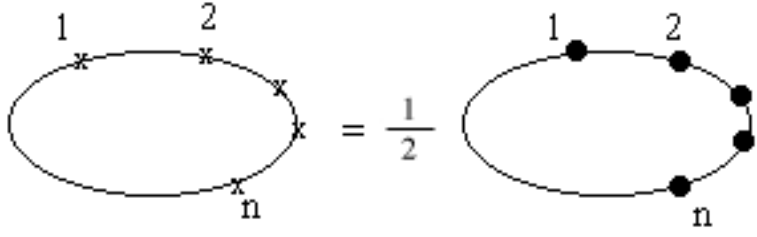}
\caption{Equation (\ref{eq:35}) in diagram form.}
\label{fig:eqn35}
\end{figure}

Equations (\ref{eq:4}) and (\ref{YMpotential}), with $t^{a}_{\mu}=0$, give
\beq
\mathcal{O} = (\Box^{2})^{2} + V_{\mu}\Box^{2}\p_{\mu}
 + W_{\mu \nu}\p_{\mu}\p_{\nu} + T_{\mu}\p_{\mu}
\label{eq:31a},
\eeq
where
\begin{eqnarray}
V^{ab}_{\mu} & = & -2 \e^{abc}\tilde{A}^{c}_{\mu},
\label{eq:31b} \\
W^{ab}_{\mu \nu} & = & -\d^{ab}\tilde{A}^{c}_{\mu}\tilde{A}^{c}_{\nu}
 + \tilde{A}^{a}_{\mu}\tilde{A}^{b}_{\nu}
 - 2 \e^{abc} \p_{\mu} \tilde{A}^{c}_{\nu},
\label{eq:31c} \\
T^{ab}_{\mu} & = & -\d^{ab}\tilde{A}^{c}_{\nu}\p_{\nu}\tilde{A}^{c}_{\mu}
 + \p_{\nu}\tilde{A}^{a}_{\mu}\tilde{A}^{b}_{\nu}
 - \e^{abc}\Box^{2}\tilde{A}^{c}_{\mu}
\label{eq:31d}.
\end{eqnarray}
Equations (\ref{eq:31b},\ref{eq:31c},\ref{eq:31d})
enable us to write
\beq
\tilde{\mathcal{O}} = (\Box^{2} + N\cdot\p)^{2},
\label{eq:32a}
\eeq
where
\beq
N^{ab}_{\mu} = -\e^{abc}\tilde{A}^{c}_{\mu}.
\label{eq:32b}
\eeq
Since the determinant is invariant under a similarity transformation,
\beq
det\;(\Box^{2} + N\cdot\p) =
 det\;(J^{-1}(\Box^{2} + N\cdot\p )J) =  det\;(\Box^{2} + Y\cdot\p + 
 P),
\label{eq:33a}
\eeq
where
\begin{eqnarray}
Y_{\mu} & = & 2 J^{-1}\p_{\mu} J + J^{-1}N_{\mu}J,
\label{eq:33b}\\
P & = & J^{-1}\Box^{2}J + J^{-1}N_{\mu}\p_{\mu}J
\label{eq:33c}.
\end{eqnarray}
Using equations (\ref{eq:33a},\ref{eq:33b},\ref{eq:33c}), equation
(\ref{eq:30}) is true if we can find a $J$ such that
\begin{eqnarray}
tr\:ln(\mathcal{I} + M\cdot\p + R) & = & 
tr\:ln
 \left(
  \mathcal{I} + \frac{2}{\Box^{2}}(Y\cdot\p + P)
 \right.
\nonumber \\[-2mm]
 & & \label{eq:34} \\[-2mm]
 & + & 
 \left. 
 \frac{1}{\Box^{2}}(Y\cdot\p + P)\frac{1}{\Box^{2}}(Y\cdot\p + P)
 \right)
\nonumber,
\end{eqnarray}
where the trace is over space-time and isospin.
Using the expansion for 
$ln(1+x)$, we find that equation (\ref{eq:34}) is equivalent to a set of
infinite, global equations,
\beq
tr\;\left\{
 \left[
  \frac{1}{\Box^{2}}(Y\cdot\p + P)
 \right]^{n}
\right\} = 
\frac{1}{2}tr\;\left\{
 \left[
  \frac{1}{\Box^{2}}(M\cdot\p + R)
 \right]^{n}
\right\}
\label{eq:35},
\eeq
with $n=1,2,\ldots,\infty$, for the nine elements of $J$ for each
space-time point.
Equation (\ref{eq:35}) in diagram form can be written as in Fig %
\ref{fig:eqn35}
where the insertions on the left are \(M\cdot\p + P\)
while those on the right are \(M\cdot\p + R\).
Because of the \((1+\fdotf)\) factors in the denominators
of the insertions, each term is essentially non-perturbative.

Can $J$ be determined by the infinite number of global equations given by
equation (\ref{eq:35})?
Naively, the answer is yes because we have infinite equations for the
infinite unknowns ($9$ components at each space-time point). 
We can visualize the problem better if we discretize 
space-time (lattice formulation).
The set of equations given by (\ref{eq:35}) becomes a set of non-linear
algebraic equations for the values of $J^{ab}(x_{i})$.

By choosing a suitable ansatz for $J$, we can always have the number of
equations equal to the number of unknowns.
It should be noted that both sides have divergencies arising from the same
point limit of the Green function of $\Box^{2}$.
This must be handled carefully using a suitable regulator.
In the continuum, what can be done is to solve for a $J$ for $n=1$ and then
verify the relationship for $n=2,3,\ldots$.
For n=1, equation (\ref{eq:35}) can be solved by the local equations
\begin{eqnarray}
Y^{aa}_{\mu} & = & \frac{1}{2}M^{aa}_{\mu}
 \label{eq:36a}, \\
R^{aa} & = & P^{aa}
 \label{eq:36b}.
\end{eqnarray}
Using equations (\ref{eq:33b},\ref{eq:33c}), and the fact that
\(N^{aa}_{\mu} = 0\),
we find
\beq
J^{-1ab} \p_{\mu}J^{ba} = \afourth M^{aa}_{\mu}
\label{eq:37a},
\eeq
\beq
J^{-1ab}\Box^{2}J^{ba} + J^{-1ab}N^{bc}_{\mu}\p_{\mu}J^{ca}
 = \half R^{aa}
\label{eq:37b}.
\eeq
There are many possible solutions to equations (\ref{eq:37a},\ref{eq:37b}).
Assume the simplest ansatz
\beq
J^{ab} = \alpha (f) \d^{ab} + \beta (f) f^{a} f^{b}
\label{eq:38},
\eeq
With \(f = (\fdotf)^{\half}\) the inverse exists under a very broad
condition, \ie, \(\alpha^{2}(\alpha + \beta\fdotf) \neq 0\).
The inverse is given by \(J^{-1ab} = \left( \frac{1}{\alpha} \right)
\d^{ab} - \left( \frac{\beta}{\alpha}\right) \frac{1}{\alpha + \beta f^{2}} 
f^{a} f^{b}\).

Equation (\ref{eq:37b}) leads to a first-order, non-linear, ordinary
differential equation for $\alpha$ and $\beta$ by equating coefficients of
\(f^{a} \p_{\mu} f^{a}\), which is the only possible structure of
$M^{aa}_{\mu}$ and $Y^{aa}_{\mu}$.
Equation (\ref{eq:37b}), on the other hand, leads to two equations -- a
second-order, non-linear ordinary differential equation by equating
coefficients of \(f^{a}\p_{\mu}f^{a}f^{b}\p_{\mu}f^{b}\) and a first-order,
non-linear ordinary differential equation by equating coefficients of
$\p_{\mu}f^{a}\p_{\mu}f^{a}$.
Seemingly, we have an overspecified problem.
Fortunately, the two first-order equations have the same structure and they
only result in the following linear relationship between $\alpha$ and
$\beta$
\beq
\alpha (f) = \mathcal{K}(f)\beta (f)
\label{eq:39a},
\eeq
where
\begin{eqnarray}
\mathcal{K} & = & 
 \frac{-f^{2}a(f) \pm \sqrt{a^{2}(f) f^{4} + 8f^{2}(f^{4} + 3f^{2} +%
 2)a(f)}}{2a(f)}
\label{eq:39b}, \\
a(f) & = &
2f^{4} + 5f^{2} - 5
\label{eq:39c}.
\end{eqnarray}
We already see from equation (\ref{eq:39b}) the multiplicity of
solutions even for the simpliest ansatz for $J$.
 
Substituting (\ref{eq:39a}) in the second-order equation, %
the problem simplifies tremendously, the differential equation becomes
linear given by
\beq
\frac{d^{2}\beta}{df^{2}} + \mathcal{B}(f)\beta + \mathcal{C}(f)\beta = 0
\label{eq:40a},
\eeq
where
\begin{eqnarray}
\mathcal{B}(f) & = &
 \frac{4}{3}\frac{f^{2}}{\mathcal{K} + f^{2}} +
 \frac{2\mathcal{K}'}{\mathcal{K}} - \frac{2}{3}\frac{f^{2}\mathcal{K}'}%
 {\mathcal{K}(\mathcal{K} + f^{2})} - \frac{1}{f(1+f^{2})}
\label{eq:40b}, \\
\mathcal{C}(f) & = &
 \frac{1}{3}\frac{\mathcal{K}''(3\mathcal{K} +
 f^{2})}{\mathcal{K}(\mathcal{K} + f^{2})} -
 \frac{\mathcal{K}'(3\mathcal{K} + f^{2})}{3f(1 + f^{2})\mathcal{K}
 (\mathcal{K} + f^{2})} 
\nonumber \\[-1mm]
 & + & \frac{(2\mathcal{K} - f^{2} - 2)f^{2}}{3\mathcal{K}(1 +
 f^{2})(\mathcal{K} + f^{2})} - \frac{(1 - f^{2})f^{2}}{3(1 + f^{2})}
\label{eq:40c}.
\end{eqnarray}
Equations (\ref{eq:40a},\ref{eq:40b},\ref{eq:40c}) are too complicated to
have a closed form solution.
However, a power series solution can be given because there are 
no poles in $f$.
Remember that in the non-linear regime \(f\neq 0\) and that $f$ is always
real.
This verifies the existence of a $J$ that at least satisfies the 
one insertion condition of the $trace$ $ln$.
Note that even at the one insertion level, the ``equality'' of the
determinant is already non-perturbative.

Finally, we answer the question, ``Why not prove equation (\ref{eq:30})
directly using $N_{\mu}$?''
The answer is simple, $\X^{aa}_{\mu}=0$ thus we cannot have an 
insertion per insertion comparison of the $tr$ $ln$s.
The single insertion trace of \(Y\cdot\p + P\) is contained in the higher
insertion traces of $N\cdot\p$ and comparison is difficult to make (but
the terms are there but with different coefficients).

\section{Proof of dimensional reduction}
\label{section:IV}

$\;\;\;\;$
The path-integral given by (\ref{eq:19a}) can be written as
\beq
\label{
eq:41a}
 PI(f;w;\psibar^{a};\psi^{a}) = 
  \int (df^{a})(dw^{a})(d\psibar^{a})(d\psi^{a}) e^{-A},
\eeq
where
\beq
\label{eq:41b}
 A = \int d^{4}x \left\{
  -\frac{1}{2}w^{2}_{a} + w_{a}\frac{\d S}{\d f^{a}}+
  \psibar^{a} \frac{\d^{2}S}{\d f^{a} \d f^{b}} \psi^{b}\right\}.
\eeq
Because of the metric $\eta^{ab}$, this action could not be derived from
the supersymmetric version of $S$, \ie,
\beq
\label{eq:42a}
 A \neq \int d^{4} d\thetabar d\theta\; \frac{1}{2}\eta_{ab}(\Phi)%
  (\p_{\mu}\Phi^{a}\p_{\mu}\Phi^{b} +%
  \p_{\theta}\Phi^{a}\p_{\thetabar}\Phi^{b}),
\eeq
where
\beq
\label{eq:42b}
 \Phi^{a} = f^{a} + \thetabar \psi^{a} + \psibar^{a}\theta +
  \thetabar\theta w^{a}.
\eeq
Also because of $\eta$, the supersymmetrized $S(\Phi)$ is not invariant
under the following transformation that leaves \(x^{2} + \thetabar\theta\)
invariant,
\begin{eqnarray}
 x_{\mu}  \rightarrow  x'_{\mu} & = & x_{\mu} + \vare_{\mu}\rhobar\theta
 + \vare_{\mu}\thetabar\rho,
 \label{eq:43a} \\
 \theta  \rightarrow  \theta' & = & \theta - 2\rho \vare \cdot x,
 \label{eq:43b} \\
 \thetabar \rightarrow \thetabar' & = & \thetabar - 2\rhobar\vare \cdot x.
 \label{eq:43c}
\end{eqnarray}
In equations (\ref{eq:43a},\ref{eq:43b},\ref{eq:43c}),
$\vare_{\mu},\rho,\rhobar$
are infinetisimal coordinate and grassman parameters.
Dimensional reduction could not be proven using the non-linear action of
$O(1,3)$.

Actually, the paper of Parisi and Sourlas hints that we should use instead
\beq
\label{eq:44}
 S_{\sigma} = \int d^{4}x \{\frac{1}{2}\p_{\mu}\sigma\p_{\mu}\sigma -
  \frac{1}{2}\p_{\mu}f^{a}\p_{\mu}f^{a} + \lambda(\sigma^{2} - \fdotf
  - 1)\}.
\eeq
The equation of motion for the multiplier $\lambda(x)$ implements the
equivalence of $S_{\sigma}$ to $S$.
We will show that the path-integral given by equation (\ref{eq:19a}) is
equal to
\begin{eqnarray}
 PI(f,\sigma,\lambda) & = & \int (df^{a})(d\sigma)(d\lambda)%
  \d \left(\frac{\d S_{\sigma}}{\d \lambda}\right)%
  \d \left(\frac{\d S_{\sigma}}{\d \sigma}\right)\;%
  \nonumber \\[-2mm]%
  &  &   \label{eq:45} \\[-2mm]%
  & \times &
  det\left(\frac{\d^{2}S_{\sigma}}{\d\phi_{i}\d\phi_{j}}\right)%
  exp\left\{-\frac{1}{2}\int d^{4}x\left(%
  \frac{\d S_{\sigma}}{\d f^{a}}\right)^{2}\right\},
  \nonumber
\end{eqnarray}
where \(\phi_{i} = (f^{a}, \sigma, \lambda)\).
The field derivatives are
\begin{eqnarray}
 \frac{\d S_{\sigma}}{\d f^{a}} & = &%
  \Box^{2}f^{a} - 2\lambda f^{a},
  \label{eq:46a} \\%
 \frac{\d S_{\sigma}}{\d\sigma} & = &%
  -\Box^{2}\sigma + 2\lambda\sigma,
  \label{eq:46b} \\%
 \frac{\d S_{\sigma}}{\d\lambda} & = &%
  \sigma^{2} - (1+\fdotf).
  \label{eq:46c}
\end{eqnarray}
It is straightforward to show that
\beq
\label{eq:47}
 \left. \frac{\d S_{\sigma}}{\d f^{a}}\right|_{\frac{d
  S_{\sigma}}{\d\lambda} = \frac{\d S_{\sigma}}{\d\sigma} = 0}%
  = \frac{\d S}{\d f^{a}}\; .
\eeq

We will now give a ``physicist's'' proof of
\beq
\label{eq:48}
 \left.
 det\left(\frac{\d^{2}S_{\sigma}}{\d\phi_{i} \d\phi_{j}}\right)%
 \right|_{\frac{\d S_{\sigma}}{\d\sigma} = \frac{\d S_{\sigma}}{\d
 \lambda} = 0}%
 \approx det\left(\frac{\d^{2}S}{\d f^{a}\d f^{b}}\right),
\eeq
where the proportionality factor is anything as long as it is
field-independent.
Note that the matrix at the LHS of (\ref{eq:48}) is an ``infinite
\(5\times 5\)'' matrix while the RHS is an ``infinite \(3\times 3\)''
matrix.
However, direct inspection shows both determinants have leading terms
$(\Box^{2})^{3}$, thus equation (\ref{eq:48}) is not really surprising.
Equation (\ref{eq:47}) can be written as
\[
 \frac{\d S}{\d f^{b}} = \left(\frac{\d S_{\sigma}}{\d \phi_{j}}\;
 \frac{\d \phi_{j}}{\d f^{b}}\right)_{\frac{\d S_{\sigma}}{\d\lambda}
 = \frac{\d S_{\sigma}}{\d \sigma} = 0}.
\]
Differentiating again with $f^{a}$, we get
\[
 \frac{\d^{2}S}{\d f^{a}\d f^{b}} = 
 \left[
  \frac{\d\phi_{i}}{\d f^{a}}
  \left(
   \frac{\d^{2}S_{\sigma}}{\d\phi_{i}\d\phi_{j}}
  \right) \frac{\d\phi_{i}}{\d f^{b}}
 \right]_{\frac{\d S_{\sigma}}{\d\lambda} = \frac{\d 
 S_{\sigma}}{\d\sigma} = 0}\; .
\]
The first matrix of the RHS of this equation is ``\(3\times 5\)
(infinite)'', the second is ``\(5 \times 5\)'', while the last is ``\(5 
\times 3\)'' for consistency with the LHS which is ``\(3 \times 3\)''.
Taking the determinant of both sides and using the cyclic property of the
determinant, we need to evaluate the determinant of the ``\(5 \times 5\) 
infinite'' matrix
\[
\left[
\begin{array}{ccc}
 1 & 0 & 0 \\[1.5mm]
 0 & 1 & 0 \\[1.5mm]
 0 & 0 & 1 \\[1.5mm]
 \frac{f^{1}}{\sigma} & \frac{f^{2}}{\sigma} & \frac{f^{3}}{\sigma}
 \\[1.5mm]
 \frac{\d\lambda}{\d f^{1}} & \frac{\d\lambda}{\d f^{2}} &
  \frac{\d\lambda}{\d f^{3}}
\end{array}
\right]
\times
\left[
\begin{array}{ccccc}
 1 & 0 & 0 & \frac{f^{1}}{\sigma} & \frac{\d\lambda}{\d f^{1}} \\[1.5mm]
 0 & 1 & 0 & \frac{f^{2}}{\sigma} & \frac{\d\lambda}{\d f^{2}} \\[1.5mm]
 0 & 0 & 1 & \frac{f^{3}}{\sigma} & \frac{\d\lambda}{\d f^{3}}
\end{array}
\right]
\]
\[
 =
\left[
\begin{array}{ccccc}
 1 & 0 & 0 & \frac{f^{1}}{\sigma} & \frac{\d\lambda}{\d f^{1}} \\[1.5mm]
 0 & 1 & 0 & \frac{f^{2}}{\sigma} & \frac{\d\lambda}{\d f^{2}} \\[1.5mm]
 0 & 0 & 1 & \frac{f^{3}}{\sigma} & \frac{\d\lambda}{\d f^{3}} \\[1.5mm]
 \frac{f^{1}}{\sigma} & \frac{f^{2}}{\sigma} & \frac{f^{3}}{\sigma} &
  \frac{\fdotf}{\sigma^{2}} & \frac{1}{\sigma}
  \left(
   \fvector\cdot\frac{\d\lambda}{\d\fvector}
  \right) \\[1.5mm]
 \frac{\d\lambda}{\d f^{1}} & \frac{\d\lambda}{\d f^{2}} &
  \frac{\d\lambda}{\d f^{3}} & \frac{1}{\sigma}
  \left(
   \fvector\cdot\frac{\d\lambda}{\d\fvector}
  \right)
  & \frac{\d\lambda}{\d\fvector} \cdot
  \frac{\d\lambda}{\d\fvector}
\end{array}
\right]
\left(
\d^{4}(x - x')
\right)
\]
Above we made use of \(\sigma = 
\left(1+\fdotf\right)^{\frac{1}{2}}\), %
and \(\;\lambda = \half \left(\frac{1}{\sigma}\right) \Box^{2}\sigma\)
 \( = \half \frac{1}{\sigma^{2}} \p_{\mu}f^{a}\p_{\mu}f^{a}.\)
The important point is that by a series of manipulations, the fourth row
or the fifth row can be made zero.
Thus,
\beq
\label{eq:49}
 det\left(\frac{\d^{2}S}{\d f^{a}\d f^{b}}\right) =
 ``0 \cdot \infty " \; 
 \left. det
  \left(
   \frac{\d^{2}S_{\sigma}}{\d\phi_{i}\d\phi_{j}}
  \right)
 \right|_{\frac{\d S_{\sigma}}{\d\lambda} = \frac{\d S_{\sigma}}{\d\sigma} 
 = 0}.
\eeq
The $\infty$ comes from the $\d^{4}(x-x')$.

Effectively, we have established that the dynamics of $f^{a}$ with action
$S$ in a random field is equivalent to a constrained stochastic dynamics
of $f^{a},\:\sigma,\:\lambda$ with action $S_{\sigma}$.
That this is true follows from the fact that we can always add
\(\half\left(\frac{\d S_{\sigma}}{\d\sigma}\right)^{2}\; {\rm and}\;
\half\left(\frac{\d S_{\sigma}}{\d\lambda}\right)^{2}\) in the
exponentials of (\ref{eq:45}).

Next we complete the proof of dimensional reduction.
First, we will exponentiate \(\d\left(\frac{\d
S_{\sigma}}{\d\sigma}\right)\) by using
\begin{eqnarray}
 \int (d\varphi)\d(\varphi - \varphi_{0}) F(\varphi) & = &
 \int (d\varphi) exp\{-\half\int dx(\varphi -\varphi_{0})^{2}\} F(\varphi)
  \nonumber \\[-2mm]
  &   &  
  \label{eq:50}  \\[-2mm]
  & + & C_{2}
  \left.
   \frac{\d^{2}F(\varphi)}{\d\varphi^{2}}
  \right|_{\varphi_{0}} + \cdots
  \nonumber
\end{eqnarray}
It may seem that the use of the above approximation unnecessarily
complicates things because we could have exponentiated
\(\d\left(\frac{\d S_{\sigma}}{\d\sigma}\right)\) by
\[
 lim_{\vare\rightarrow 0}\ exp\left\{-\frac{1}{2\vare}\int
 d^{4}x\: \left(\frac{\d S_{\sigma}}{\d\sigma}\right)^{2}\right\}.
\]
Unfortunately, the presence of $\vare$ invalidates $SUSY$ except for
$\vare=1$.

Second, we will exponentiate \(\d \left(\frac{\d
S_{\sigma}}{\d\lambda}\right)\) by introducing an auxilary field $w_{x}$
by
\[
 \d\left(
  \frac{\d S_{\sigma}}{\d\lambda}
 \right) = \int (dw_{\lambda})\: e^{-\int d^{4}x w_{\lambda} \frac{\d
 S_{\lambda}}{\d\lambda}}.
\]
This was not done for \(\frac{\d S_{\sigma}}{\d\sigma}\) because $\sigma$
has a kinetic term (and $\lambda$ does not), which requires the presence of
a \(\half w^{2}_{\sigma}\) (will arise from
$\p_{\theta}\Phi_{\sigma}\p_{\thetabar}\Phi_{\sigma}$).

Taking everything into account, we have
\begin{eqnarray}
 PI(\ref{eq:18}) & = & PI(\ref{eq:19a}) \nonumber \\
 & = & PI(\ref{eq:45}) \nonumber \\
 & \approx &
 \int (df^{a})(d\sigma)(d\lambda)\: det
 \left(
  \frac{\d^{2}S_{\sigma}}{\d\phi_{i}\d\phi_{j}}
 \right) \nonumber \\
 &   &  \label{eq:51} \\
 & \times & exp
 \left\{
  -\int d^{4}x
  \left[
   +\half\left(\frac{\d S_{\sigma}}{\d f^{a}}\right)^{2}
   +\half\left(\frac{\d S_{\sigma}}{\d\sigma}\right)^{2}
   + w_{\lambda}\frac{\d S_{\lambda}}{\d\lambda}
  \right]
 \right\}, \nonumber
\end{eqnarray}
where the numbers in the $PI$ refer to the equation numbers in this paper.
Note that the last equation only involves an approximation, because we
neglected the other terms in equation (\ref{eq:50}).
Since the kinetic term of $f^{a}$ has the wrong sign, we momentarily
shift \(if^{a} \rightarrow f^{a}\), effectively rotating the $O(1,3)$
symmetry to $O(4)$.
Finally, we introduce the fermion fields for the determinant and the
auxillary fields $w_{a}$ and $w_{\sigma}$ to complete the squares of the
first two terms of the exponential to get
\beq
\label{eq:52a}
 PI(\ref{eq:45}) \approx \int (d\phi_{i}) (dw_{i}) (d\psibar_{i})
 (d\psi_{j})\: exp\{-A_{SS}\},
\eeq
where \(\phi_{i} = (f^{a}, \sigma, \lambda);\; w_{i} = (w^{a},
w_{\sigma}, w_{\lambda});\; \psi_{i} = (\psi^{a}, \psi_{\sigma},
\psi_{\lambda})\); and
\begin{eqnarray}
 A_{SS} & = & \int d^{4}x
 \left\{
  -\half w^{2}_{a} - \half w^{2}_{\lambda} + w_{a}\frac{\d
  S_{\sigma}}{\d f^{a}} + w_{\sigma} \frac{\d S_{\sigma}}{\d\sigma}
 \right.  \nonumber \\[-2mm]
 &   &  
 \label{eq:52b}  \\[-2mm]
 & + &
 \left.
  w_{\lambda} \frac{\d S_{\sigma}}{\d\lambda} + \psibar_{i}
  \frac{\d^{2}S_{\sigma}}{\d\psi_{i}\d\psi_{j}} \psi_{j}
 \right\}. \nonumber
\end{eqnarray}
Equation (\ref{eq:52b}) can be derived from the supersymmetric version of
$S_{\sigma}$, \ie,
\begin{eqnarray}
 A_{SS} = S_{\sigma}(\Phi) & = & \int d^{4}x d\thetabar d\theta
 \left\{
  \half\p_{\mu}\Phi_{\sigma}\p_{\mu}\Phi_{\sigma}
  + \half\p_{\theta}\Phi_{\sigma}\p_{\thetabar}\Phi_{\sigma}
 \right. \nonumber \\[-2mm]
 &   &
\label{eq:53}      \\[-2mm]
 & + &
 \left.
  \half\p_{\mu}\Phi^{a} \p_{\mu}\Phi^{a}
  + \half \p_{\theta}\Phi^{a} \p_{\thetabar}\Phi^{a}
  + \Phi_{\lambda} (\Phi^{2}_{\sigma} + \Phi^{a} \Phi^{a} - 1)
 \right\} \nonumber
\end{eqnarray}
where $\Phi^{a}$ is given by equation (\ref{eq:42b}) and
\begin{eqnarray}
 \Phi_{\sigma} & = & 
 \sigma + \thetabar \psi_{\sigma} + \psibar_{\sigma}\theta +
 \thetabar\theta w_{\sigma},
  \nonumber  \\
 \Phi_{\lambda} & = & \lambda + \thetabar\psi_{\lambda} +
  \psibar_{\lambda}\theta + \thetabar\theta w_{\lambda}.
  \nonumber
\end{eqnarray}
Equation (\ref{eq:53}) explicitly shows invariance under equations
(\ref{eq:43a},\ref{eq:43b},\ref{eq:43c}) and dimensional reduction
follows.
Then we rotate back \(f^{a} \rightarrow if^{a}\), yielding again the
$O(1,3)$ sigma model in the linear form but in dimension reduced by two.

This completes the proof that the non-perturbative regime of Yang-Mills
theory in $4$D is equivalent to a non-linear $O(1,3)$ sigma model in
$2$D.

\section{Conclusion}

$\;\;\;\;$
We have exhibited the Parisi-Sourlas mechanism in
Yang-Mills theory.
Since the starting point of the proof is the scalar field derived from the
non-linear gauge condition, this paper proves further that the Coulomb
gauge is not an appropriate gauge fixing in non-Abelian theories.
Important field configurations will be missed in the path-integral as
shown in reference \cite{Jamag:MPLa}, where the linear potential was
derived, and in this paper, where the Parisi-Sourlas mechanism was
exhibited.

It is also important to point out that the derivation of the linear
potential is not quantum mechanical but merely a statistical treatment of
a class of classical configurations, \ie, all spherically symmetric
$f^{a}(x)$ in $4$D Euclidean space.
But as this paper showed, taking into account the full dynamics, including
quantum effects, results in dimensional reduction and equivalence to an
$O(1,3)$ sigma model in $2$D.

\section*{Acknowledgement:}

$\;\;\;\;$
This research was supported in part by the UNESCO-Jakarta office and the
Commission on Higher Education of the Republic of the Philippines.
Part of the work was carried out when the author visited the Abdus
Salam-ICTP.
He is grateful to Seif Randjbar-Daemi for making his visit to ICTP
possible.
The author would like to thank Martin Reuter for making his stay in Mainz
possible and for the illuminating discussions.
He would also like to thank Johnrob Y. Bantang and Nelson Y. Caroy of
the National Institute of Physics, the University of the Philippines, 
for typing the manuscript in \LaTeX.


\begin{thebibliography}{9}
\bibitem{Seiberg-Witten}N. Seiberg and E. Witten, %
 {\it Nuclear Physics} {\bf B 426} (1994) 19
\bibitem{'tHooft}G. `t Hooft, %
 {\it Nuclear Physics} {\bf B 190} [FS3] (1981) 455 \\
 S. Mandelstam, {\it Phys. Reports} {\bf 23} (1976) 245
\bibitem{Fad-Niemi}L. Fadeev and A.J. Niemi, %
 {\it Phys. Rev. Let.} {\bf 82} (1999) 1624
\bibitem{Jamag:MPLa}J.A. Magpantay, %
 {\it Modern Phys. Lett. A} {\bf 14} (1999) 447
\bibitem{Jamag:PTP}J.A. Magpantay, %
 {\it Prog. of Theor. Phys.} {\bf 91} (1994) 573
\bibitem{Jamag:etc}J.A. Magpantay and E Cuasing, Jr, %
 {\it Mod. Phys. Lett.} {\bf A11} (1996) 87
\bibitem{Jamag:workshop}J.A. Magpantay, in $2^{nd}$ Jagna Int. Workshop on
 Mathematical Methods of Quantum Physics, eds. %
 C.C. Bernido, M.V. Carpio-Bernido, K. Nakamura, and K. Watanabe (Gordon
 and Breach, 1999), p. 241.
\bibitem{Gribov}V.N. Gribov, %
 {\it Nuclear Physics} {\bf B 139} (1978) 1
\bibitem{Parisi-Sourlas}G. Parisi and N. Sourlas, %
 {\it Phys. Rev. Lett.} {\bf 43} (1979) 744
\end{thebibliography}
\end{document}